# Operation of Graphene Transistors at GHz Frequencies

*Yu-Ming Lin\*, Keith A. Jenkins, Alberto Valdes-Garcia,*

*Joshua P. Small, Damon B. Farmer, and Phaedon Avouris*

Email: yming@us.ibm.com

IBM T. J. Watson Research Center, Yorktown Heights, NY 10598

ABSTRACT

Top-gated graphene transistors operating at high frequencies (GHz) have been fabricated and their characteristics analyzed. The measured intrinsic current gain shows an ideal 1/f frequency dependence, indicating an FET-like behavior for graphene transistors. The cutoff frequency $f_T$ is found to be proportional to the dc transconductance $g_m$ of the device, consistent with the relation $f_T = g_m/(2\pi C_G)$. The peak $f_T$ increases with a reduced gate length, and $f_T$ as high as 26 GHz is measured for a graphene transistor with a gate length of 150 nm. The work represents a significant step towards the realization of graphene-based electronics for high-frequency applications.

Graphene is a two-dimensional (2D) material with great potential for electronics [1, 2]. With essentially the same lattice structure as an unwrapped carbon nanotube [3], graphene shares many of the advantages of nanotubes, such as the highest intrinsic carrier mobility at room temperature of any known materials [4-6]. This makes these carbon-based electronic materials particularly promising for high-frequency circuits. However, due to the high impedance of a single carbon nanotube transistor, high-frequency properties of nanotubes were investigated indirectly using various mixing techniques [7, 8], and direct ac measurements of these devices at GHz frequencies were realized only recently [9, 10], enabled by the larger device current in nanotube arrays. In contrast, one distinct advantage of graphene lies in its 2D nature, so that the drive current of a graphene device, in principle, can be easily scaled up by increasing the device channel width. This width scaling capability of graphene is of great significance for realizing high-frequency graphene devices with sufficient drive current for large circuits and associated measurements. Furthermore, the planar graphene allows for the fabrication of graphene devices and even integrated circuits utilizing well-established planar processes in the semiconductor industry. Recently, it was shown that graphene devices can exhibit current gain in the microwave frequency range [11]. Despite intense activities on graphene research, the intrinsic high-frequency transport properties of graphene transistors have not been systematically studied.

This letter presents the first comprehensive experimental studies on the high-frequency response of top-gated graphene transistors for different gate voltages and gate lengths. The intrinsic current gain of the graphene transistors was found to decrease with increasing frequency and follows the ideal 1/f dependence expected for conventional FETs. This not only verifies the ac measurement and de-embedding procedures used here for extracting the intrinsic high-frequency properties, but also suggests a conventional FET-like behavior for graphene transistors. The cutoff frequency $f_T$ deduced from S parameter measurements exhibits strong gate voltage dependence and is proportional to the dc transconductance. The peak cut-off frequency is found to be inversely proportional to the square of the gate length, and for a gate length of 150 nm, a peak $f_T$ as high as 26 GHz is obtained.

Figure 1 shows the device layout of graphene field-effect transistors with probe pads designed for high-frequency measurements. Graphene was prepared by mechanical exfoliation on a high-resistivity Si substrate (>10 kΩ·cm) covered by a layer of 300nm thermal $SiO_2$, and Raman spectroscopy was employed to count the number of graphene layers. The inset of Fig. 1(b) shows the optical image of a graphene flake, where the region on the left was identified to be single-layer graphene. Source and drain electrodes made of 1 nm Ti as the adhesion layer and 50 nm-thick Pd were defined by e-beam lithography and lift-off. A 12-nm-thick $Al_2O_3$ layer was then deposited by atomic layer deposition (ALD) at 250ºC as the gate insulator. In order to form a uniform coating of oxide on graphene, a functionalization layer consisting of 50 cycles of $NO_2$-TMA (trimethylaluminum) was first deposited prior to the growth of gate oxide [12, 13]. This $NO_2$-TMA functionalization layer was essential for the ALD process to achieve thin (<10 nm) gate dielectrics on graphene without producing pinholes that cause gate leakage. The dielectric constant of ALD-grown $Al_2O_3$ is determined by *C-V* measurements and found to be about 7.5. Lastly, 10nm/50nm Pd/Au was deposited and patterned to form the top gate. As shown in Fig. 1(b), the source electrodes were designed to overlap the entire graphene flake (see figure inset) in order to minimize the uncertainty in the de-embedding process for high-frequency S-parameter measurements, as explained below. In the device shown in Fig. 1(b), the distance between the source and drain electrodes is 500 nm, and the top gate underlaps the source-drain gap with a gate length $L_G$ of 360 nm. The total gate width (or channel width), including both channels, is ~ 40μm. Fig. 1(a) shows the optical image of the complete device layout where the standard ground-signal-ground probing pads are realized for the gate and the drain to allow for transition from coax to on-chip coplanar waveguide (CPW) electrodes.

Measurements of dc electrical properties of graphene devices were performed in order to gain insight into their high-frequency response. In addition, the dc electrical characteristics were monitored at each fabrication step so that issues affecting the final device performance could be identified. Fig. 2 shows the conductance, $G = I_D/V_D$, of a graphene device before ALD oxide deposition, using the Si substrate as the back gate and a dc source-drain bias $V_D$ of 100 mV. The field-effect mobility $\mu_{eff}$ can be calculated using the relation $\Delta\sigma = q\,\Delta n\,\mu_{eff}$, where $n$ is the 2D carrier density that is controlled by the gate voltage. For this as-prepared graphene device, $\mu_{eff}$ is estimated to be 400 cm$^2$/V·s. After the deposition of top gate dielectrics by ALD, there was significant reduction in both the device conductance and field-effect mobility (see inset of Fig. 2)). The current and mobility degradation in the oxide-covered devices may be attributed to charged impurity scattering associated with the NO$_2$ functionalization layer and interface phonon scattering in the oxide [4]. Similar impact of the oxide environment on electrical properties of graphene was previously observed [14, 15], and these results indicate that further oxide optimization or alternative dielectrics are required in order to benefit from the high intrinsic mobilities in graphene.

The dc electrical characteristics of the completed graphene device after the deposition of the top-gate electrode are shown in Fig. 3(a). The inset shows the measured current as a function of (top-gate) voltage $V_G$ at a drain bias of $V_D$ = 100 mV. Despite the small on/off ratio, the graphene devices are essentially ambipolar field-effect transistors, as indicated by the "V"-shape gate dependence in the measured $I_D$-$V_G$ curve. In these graphene field-effect transistors (GFET), the transport is dominated by electrons and holes for positive and negative gate voltages, respectively, and the conductance minimum is denoted as the Dirac point where electrons and holes make equal contributions to the transport. Compared with the inset of Fig. 2 for the back-gated device, the top-gated GFET exhibits no discernible changes in minimum conductance or the on-current, indicating that metallization of top-gate electrodes did not alter the properties of graphene channel. Fig. 3(a) shows the *n*-type output characteristics, $I_D$-$V_D$, of the graphene transistor at various gate voltages. It is found that the top-gated GFETs studied here exhibit a nearly linear $I_D$-$V_D$ dependence up to 1.6 V for the gate voltage ranges measured. This lack of current saturation is due to the fact that graphene is a zero-gap semiconductor. It has been suggested that velocity saturation at higher biases may lead to the current saturation phenomenon in graphene transistors [16]. However, a higher carrier mobility may be required to achieve this saturation velocity within the drain bias of practical interest.

The high-frequency performance of a transistor for small-signal response is mainly determined by the transconductance $g_m = dI_D/dV_G$. Fig. 3(b) shows the measured transconductance of the GFET as a function of gate voltage $V_G$ for different drain biases. Both the sign and magnitude of $g_m$ are found to be strongly dependent on the gate voltage. The negative $g_m$ branch represents *p*-type transport dominated by holes, and $g_m$ changes sign with increasing gate voltage as the channel becomes *n* type. For both branches, the magnitude of $g_m$ rises with increasing $V_D$ because the graphene transistor possesses nearly linear $I_D$-$V_D$ output characteristics in this operation regime. At $V_D$ = 1.6 V, the peak $g_m$ values are ~45 mS/mm and -35mS/mm for *n* and *p* branches, respectively, in this 360-nm-gate device.

To probe the high-frequency response of the GFET, on-chip microwave measurements were carried out using a HP8510 vector network analyzer up to 10 GHz to obtain the scattering S-parameters that relate the ac currents and voltages between the drain and the gate of the GFET. A standard open, short and load calibration was employed to calibrate the network analyzer, and short and open structures were used to de-embed the signals of the parasitic capacitance and the series resistance associated with the pads and connections. This de-embedding procedure is a well-established standard and is performed here using a specific "open" test structure without graphene and a "short" test device where the gate, source, and drain electrodes are all connected by metals. To achieve high fidelity in the de-embedding process, the layouts of these open and short structures are strictly identical to that of the active device except in the graphene channel. The de-embedded S parameters constitute a complete set of coefficients to describe intrinsic input and output behaviors of the graphene device, and can be used to derive other important electrical properties such as gain. Fig. 4(a) shows the magnitude of all of the four de-embedded S-parameters of the GFET measured at $V_G$ = 0.5V and $V_D$ = 1.6 V as a function of frequency, and the corresponding short circuit small-signal current gain $h_{21}$ = $i_D/i_G$ calculated from the measured S parameters is shown in Fig. 4(b).

In Fig. 4(b), the de-embedded current gain $h_{21}$ decreases with increasing frequency following the 1/f slope expected for a conventional FET. In a regular FET, this 1/f frequency dependence of $h_{21}$, equivalent to a decay slope of -20dB/decade, results from the gate impedance given by $Z = 1/j\omega C_G$, where $\omega = 2\pi f$ and $C_G$ is the gate capacitance, that decreases with increasing frequency. Therefore, the 1/f dependence of current gain obtained in Fig. 4(b) is significant because it not only validates the high-frequency measurements and the de-embedding procedures used to extract the intrinsic GFET characteristics, but it also suggests regular FET-like behaviors for graphene transistors as a function of frequency. One of the important figures of merit for characterizing high-frequency transistors is the cut-off frequency $f_T$, defined as the frequency where the current gain becomes unity ($h_{21}$ = 1). In practice, for a transistor possessing the ideal -20dB/decade slope for $h_{21}$, the cut-off frequency $f_T$ is determined by the product of $h_{21}$ and frequency, i.e. $f \times h_{21}(f)$, over the measured frequency range. Thus, for the device shown in Fig. 4, the cut-off frequency $f_T$ can be determined by either approach to be ~ 4 GHz.

The high-frequency operation of the graphene transistor is found to be highly dependent on the dc bias condition. Fig. 5 shows the measured cut-off frequency $f_T$ of the GFET as a function of gate voltage. At all gate voltages, the de-embedded current gain $h_{21}$ exhibits the 1/f frequency dependence similar to that shown in Fig. 4(b), so that the cut-off frequency can be reliably determined. The *n*-branch of the graphene transistor is shown here because of the higher transconductance for electrons than for holes in this device. The dc transconductance $g_m$ of the GFET device is also plotted in Fig. 5 (right axis) as a function of gate voltage, where the strong correlation between $g_m$ and $f_T$ can be clearly observed. As the gate voltage is varied, $f_T$ follows the modulation of $g_m$, with a maximum cut-off frequency of ~ 4 GHz corresponding to the peak transconductance of 1.6 mS at $V_G = 0.5$ V. The inset of Fig. 5 plots the measured $f_T$ as a function of device transconductance $g_m$, showing a linear dependence between the two quantities. In an FET, the cut-off frequency is proportional to the transconductance and given by $f_T = g_m/(2\pi C_G)$, where $C_G$ is the total gate capacitance. Using this relation, a gate capacitance of $C_G \cong 72$ fF can be extracted by the fitted slope in the inset of Fig. 5. This gate capacitance can also be independently estimated from the device geometry and gate dielectrics. Based on the gated area of approximately 360nm by 40 μm for this GFET, the gate capacitance is calculated to be ~80 fF, which is in close agreement with extracted values from $f_T$ and $g_m$ measurements. These results show that the high-frequency behavior of these graphene transistors can be described as an FET with a static, constant gate capacitance within a significant portion of the bias range.

In principle, the maximum cut-off frequency of an FET can be improved by reducing the gate length. To investigate the length dependence of $f_T$ in graphene devices, graphene transistors with various gate lengths down to 150 nm were fabricated and investigated for their high-frequency operations. All of the graphene devices studied here were prepared in one batch and on the same chip in order to minimize the device-to-device variations introduced in the fabrication processes. As before, mobility degradation was observed in all devices after ALD oxide deposition. The maximum $f_T$ was found to increase with reduced gate lengths, as expected, and for the 150-nm-gate GFET, a peak cut-off frequency as high as 26 GHz was obtained, as shown in Fig. 6. To the authors' knowledge, this is the highest value measured for graphene transistors to date. The inset of Fig. 6 shows the maximum cut-off frequency as a function of gate length $L_G$ measured at a drain bias $V_D = 1.6$ V. By reducing the gate length from 500 to 150 nm, the maximum $f_T$ was increased from 3 to 26 GHz. Furthermore, the length dependence of maximum $f_T$ in graphene transistors can be fit by $f_T \sim 1/L_G^2$, as shown by the solid line in the inset of Fig. 6. This $1/L_G^2$ length dependence of maximum $f_T$ can be qualitatively understood based on the transient time of carriers in the graphene channel, as follows. The cut-frequency $f_T$ of a transistor is mainly determined by the minimum time $\tau$ required for a carrier to travel across the channel, i.e. $f_T \propto 1/\tau = v_d/L_G$, where $v_d$ is the carrier drift velocity. Since the devices considered here operate in the linear $I_d$-$V_d$ regime, the drift velocity is proportional to the drive field given by $v_d = \mu_{eff} \cdot E_d$, where $E_d$ is the electric field in the channel that is inversely proportional to the gate length for a given drain bias. Therefore, it follows that $f_T \sim \mu_{eff} \cdot (V_D/L_G)/L_G \sim 1/L_G^2$.

In summary, top-gated graphene transistors of various gate lengths were fabricated and their high-frequency response was directly characterized by standard S-parameter measurements. The short-circuit current gain showed the ideal $1/f$ frequency dependence, confirming the measurement quality and the FET-like behavior for graphene devices. As the gate voltage is varied, the measured $f_T$ was found to be proportional to the dc transconductance $g_m$, following the relation $f_T = g_m/(2\pi\, C_G)$. Furthermore, $f_T$ was found to increase with decreasing channel length, with the scaling dependence $f_T \sim 1/L_G^2$ for the GFETs studied here. A peak cut-off frequency $f_T$ as high as 26 GHz was measured for a 150-nm-gate graphene transistor, establishing the state of the art for graphene transistors. These results also indicate that if the high mobility of graphene can be preserved during the device fabrication process, a cut-off frequency approaching THz may be achieved for graphene FET with a gate length of just 50nm and carrier mobility of 2000 cm$^2$/V·s.

The authors would like to thank F. Xia and C. Y. Sung for fruitful discussions on device fabrication and J. Tsang for Raman scattering characterization. They also thank Bruce Ek, M. Rooks, and J. Bucchignano for the expert technical assistance. This work is supported by DARPA under contract FA8650-08-C-7838 through the CERA program.

FIGURE CAPTIONS

Fig. 1: (a) Optical image of the device layout with ground-signal-ground accesses for the drain and the gate. (b) (False color) SEM image of the graphene channel and contacts. The inset shows the optical image of the as-deposited graphene flake (circled area) prior to the formation of electrodes. (c) Schematic cross-section of the graphene transistor. Note that the device consists of two parallel channels controlled by a single gate in order to increase the drive current and device transconductance.

Fig. 2: Measured conductance as a function of back-gate voltage $V_{BG}$ of the graphene transistor before depositing the top-gate dielectric. The inset shows the same device after the deposition of $Al_2O_3$ by ALD. The two arrows represent the sweeping direction of the gate voltage.

Fig. 3: (a) Measured output characteristics of the graphene transistor for various top-gate voltages. The inset shows the transfer characteristics at a drain voltage of 100 mV. (b) Measured transconductance as a function of gate voltage $V_G$ for different drain voltages. The transconductance changes sign as gate voltages increases because of the ambipolar transport. For each drain voltage, the magnitude of the transconductance exhibits peak values at negative and positive gate voltages, corresponding to the *p*- and *n*-branch of the carrier transport, respectively.

Fig. 4: (a) Measured scattering parameters, $S_{11}$, $S_{12}$, $S_{21}$ and $S_{22}$, of a graphene transistor after de-embedding. (b) The current gain $h_{21}$ calculated from the measured S parameters as a function of frequency. The de-embedded current gain $h_{21}$ is inversely proportional to the frequency for a significant portion of the frequency range. The solid line corresponds to the ideal 1/f dependence, or equivalently, -20 dB/decade slope, of the current gain. The cut-off frequency is determined to be 4 GHz. The drain and gate voltages are 1.6 V and 0.5 V, respectively.

Fig. 5: Measured cut-off frequency $f_T$ (left axis) and dc transconductance $g_m$ (right axis) as a function of gate voltage. The drain bias is 1.6 V. Inset: measured $f_T$ as a function of $g_m$, showing the linear dependence between the two quantities.

Fig. 6: Measured current gain $h_{21}$ as a function of frequency of a GFET with $L_G$=150 nm, showing a cut-off frequency at 26 GHz. The dashed line corresponds to the ideal 1/f dependence for $h_{21}$. Inset: the maximum $f_T$ as a function gate length for the four GFETs measured. All devices are biased at $V_D$ = 1.6 V. The solid line corresponds to the $1/L_G^2$ dependence for peak $f_T$.

REFERENCES

[1] Y. Zhang, J. W. Tan, H. L. Stormer, and P. Kim, Nature 438, 201 (2005).

[2] A. K. Geim and K. S. Novoselov, Nature Mater. 6, 183 (2007).

[3] Ph. Avouris, Z. Chen, and V. Perebeinos, Nature Nanotechnology 2, 605 (2007).

[4] J. H. Chen, C. Jang, S. Xiao, M. Ishigami, and M. S. Fuhrer, Nature Nanotechnology 3, 206 (2008).

[5] K. I. Bolotin, K. J. Sikes, Z. Jiang, M. Klima, G. Fudenberg, J. Hone, P. Kim, and H. L. Stormer, Solid State Communications 146, 351-355 (2008) 146, 351 (2008).

[6] S. Morozov, K. Novoselov, M. Katsnelson, F. Schedin, D. Elias, J. Jaszczak, and A. Geim, Phys. Rev. Lett. 100, 016602 (2008).

[7] J. Appenzeller and D. J. Frank, Appl. Phys. Lett. 84, 1771 (2004).

[8] S. Rosenblatt, H. Lin, V. Sazoneva, S. Tiwari, and P. L. McEuen, Appl. Phys. Lett. 87, 153111 (2005).

[9] A. L. Louarn, F. Kapche, J.-M. Bethoux, H. Happy, G. Dambrine, V. Derycke, P. Chenevier, N. Izard, M. F. Goffman, and J.-P. Bourgoin, Appl. Phys. Lett. 90, 233108 (2007).

[10] C. Rutherglen, D. Jain, and P. Burke, Appl. Phys. Lett. 93, 083119 (2008).

[11] I. Meric, N. Baklitskaya, P. Kim, and K. Shepard, IEDM Technical Digest, in press (2008).

[12] D. B. Farmer and R. G. Gordon, Nano Lett. 6, 699 (2006).
12

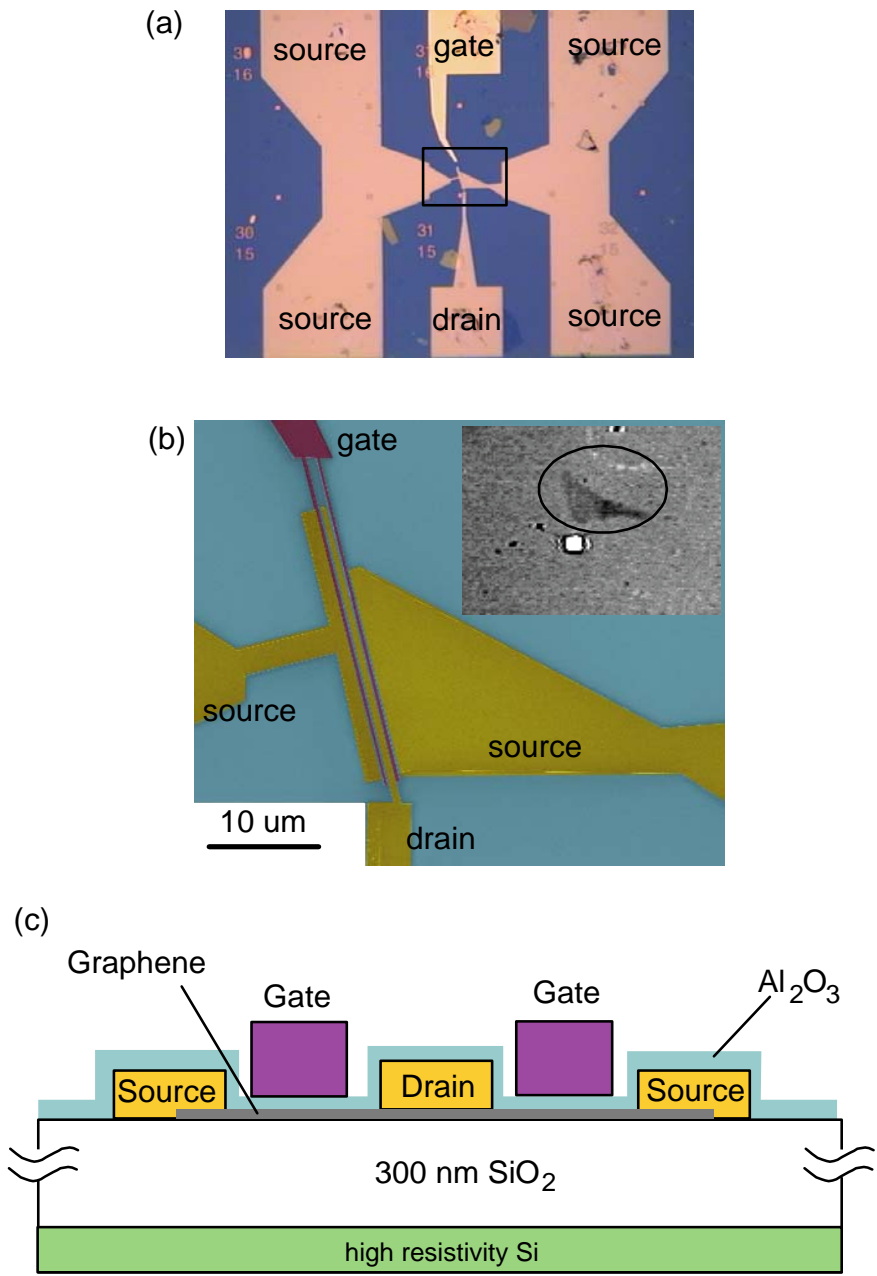

Fig. 1: Lin et al.

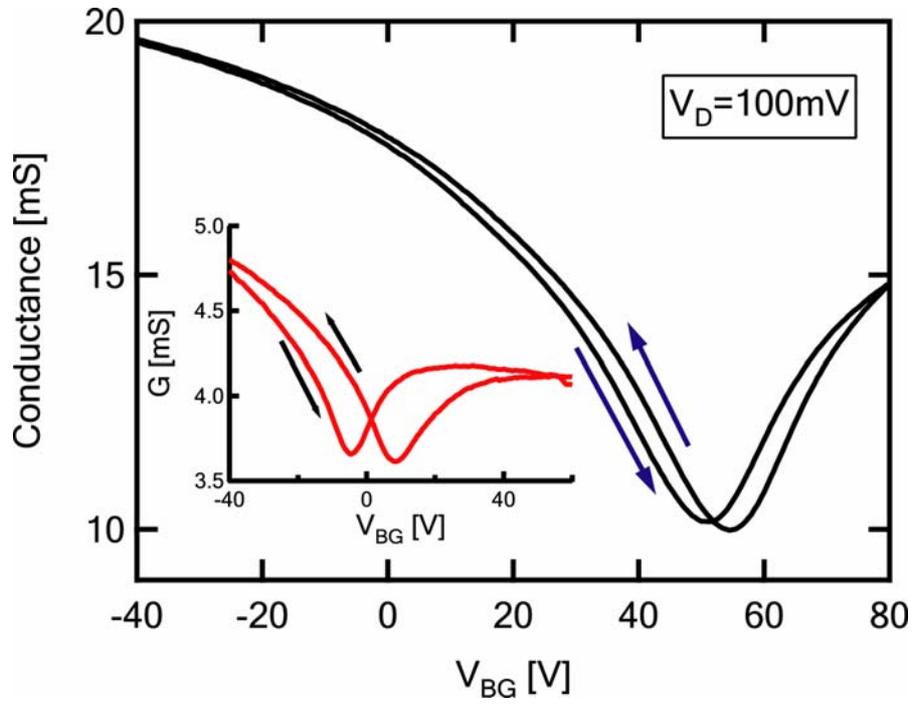

Fig. 2: Lin et al.

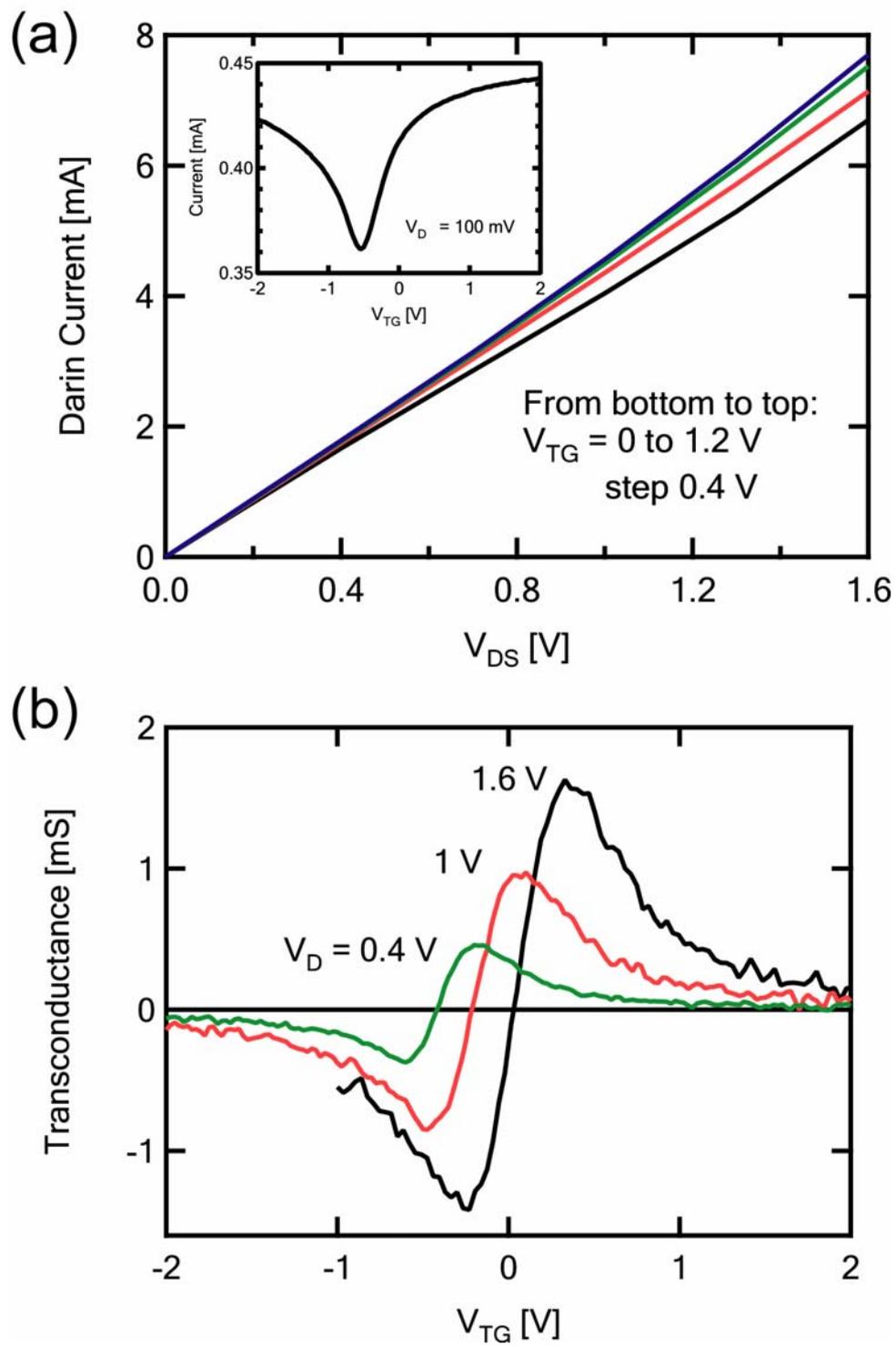

Fig 3: Lin et al.

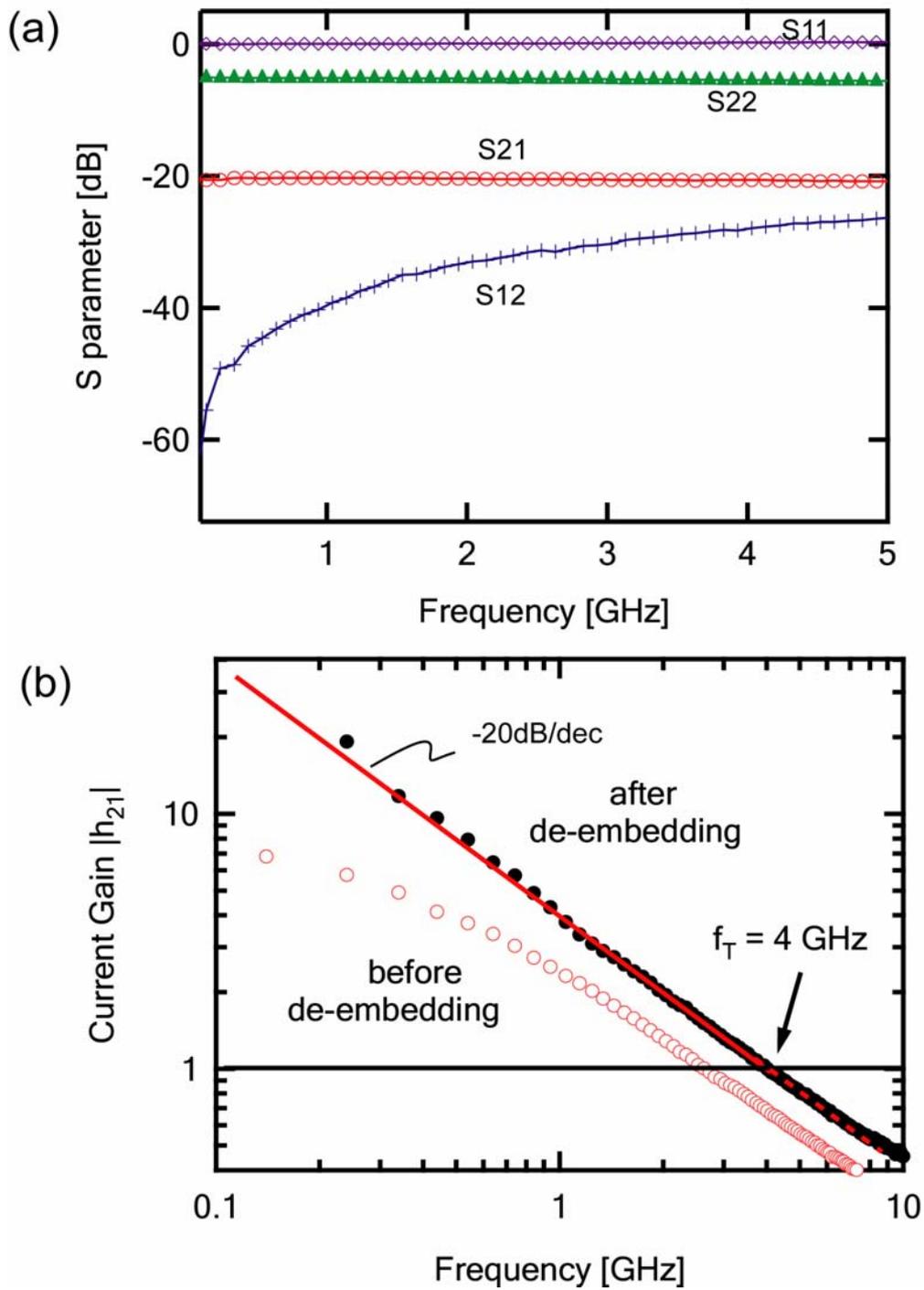

Fig 4: Lin et al.

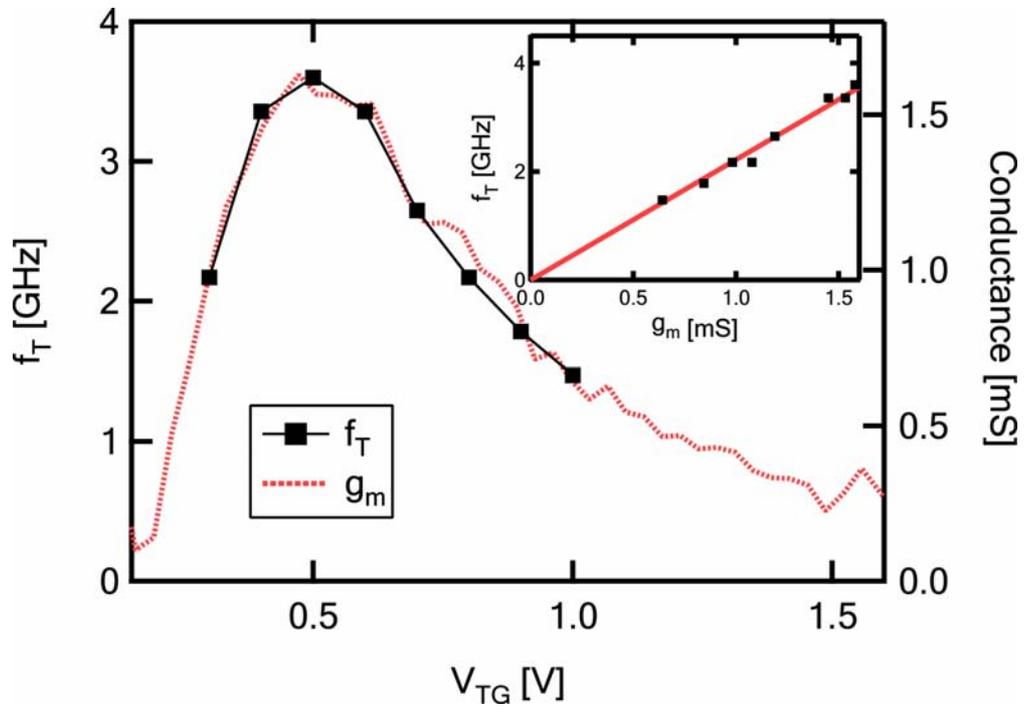

Fig. 5: Lin et al.

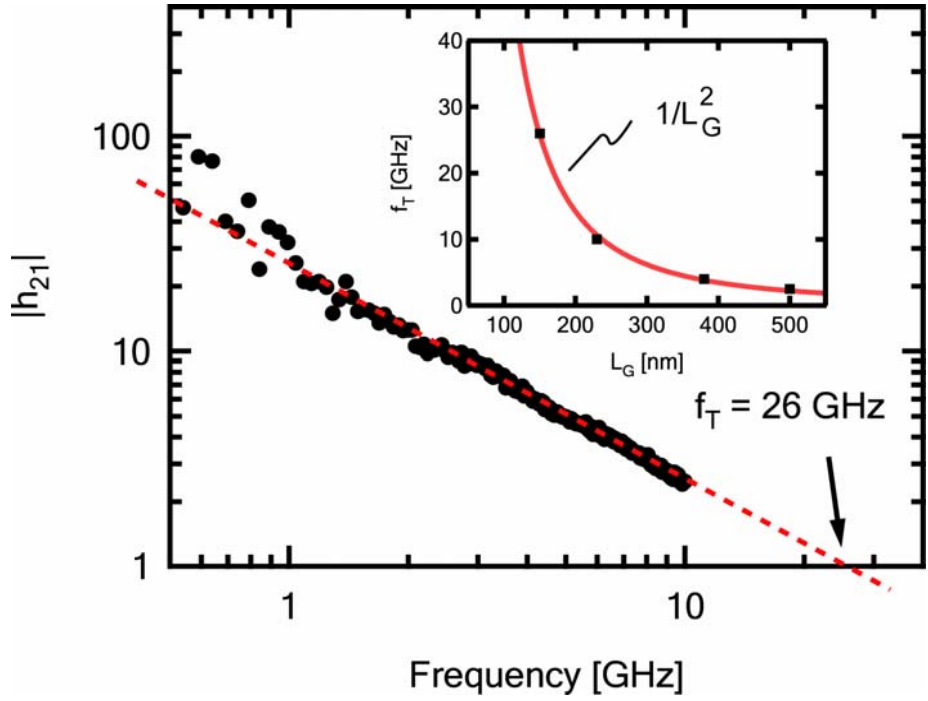

Fig. 6: Lin et al.